\documentclass[letterpaper, 10 pt, conference]{ieeeconf}  

\IEEEoverridecommandlockouts                              

\usepackage{graphics} 
\usepackage{epsfig} 
\usepackage{mathptmx} 
\usepackage{times} 
\usepackage{amsmath} 
\usepackage{amssymb}  
\usepackage{stmaryrd}
\usepackage{hyperref}
\usepackage{subcaption}
\usepackage{thmtools}
\usepackage{multirow}
\usepackage[ruled,vlined]{algorithm2e}

\usepackage[pdftex,dvipsnames]{xcolor}
\usepackage[color=SkyBlue!50]{todonotes}

\usepackage{fainekos-macros}
\usepackage{rod-macros}

\title{\LARGE \bf
Combined Left and Right Temporal Robustness \\for Control under STL Specifications
}

\author{Al\"ena Rodionova, Lars Lindemann, Manfred Morari and 
	George J. Pappas$^\dagger$
\thanks{$^\dagger$ The authors are with the Department of Electrical and Systems Engineering, University of Pennsylvania, Philadelphia PA, USA. 
  {\tt\small\{nellro,\, larsl,\, morari,\, pappasg\}@seas.upenn.edu}.}%
\thanks{This work was supported by the AFOSR under grant FA9550-19-1-0265 Assured Autonomy in Contested Environments.}
}

\begin{document}

\maketitle
\thispagestyle{empty}
\pagestyle{empty}

\begin{abstract}
Many modern autonomous systems, particularly multi-agent systems, are time-critical and need to be robust against timing uncertainties. Previous works have studied  left and right time robustness of signal temporal logic specifications by considering time shifts in the predicates that are either only to the left or only to the right.  We propose a combined notion of temporal robustness which simultaneously considers  left and right time shifts.  \edit{For instance, in a scenario where a robot plans a trajectory around a pedestrian, this combined notion  can now capture uncertainty of the pedestrian arriving earlier or later than anticipated.}  We first derive  desirable properties of this new notion with respect to  left and right time shifts and then design control laws for linear systems that maximize temporal robustness using mixed-integer linear programming. Finally, we present two case studies to illustrate how the proposed temporal robustness accounts for timing uncertainties.

\vspace{10pt}
\end{abstract}

\section{INTRODUCTION}

This paper studies temporal robustness of time-critical systems, i.e., systems in which meeting real-time safety constraints is of great importance. Examples of time-critical systems include multi-robot systems and self-driving cars. While time-critical systems may satisfy their safety constraints under nominal operating conditions, already slight temporal perturbations such as time delays may jeopardize its safety if the system is not robust against such perturbations. 

A common way to express real-time constraints is to use signal temporal logic (STL) \cite{maler2004monitoring}. Spatial robustness of STL specifications, quantifying permissible spatial perturbations, has been widely studied in the literature, see e.g.,  \cite{fainekos2009robustness,gilpin2020smooth,varnai2020robustness}. For control under spatial robustness objectives, there exist mixed-integer linear programming (MILP) approaches \cite{raman2014model,buyukkocak2021planning,kurtz2022mixed}, gradient-descent  searches \cite{mehdipour2019average,pant2018fly}, control barrier functions for STL \cite{lindemann2018control,charitidou2021barrier}, and learning-based frameworks \cite{cai2022overcoming}. However, these notions do not directly capture any robustness against temporal uncertainties. A first attempt to define time robustness for STL specifications was made in \cite{Donze10STLRob}. The authors define the left (right) time robustness by quantifying the maximal permissible left (right) time shifts in the predicates of the STL specification that do not result in a violation of the  specification. In our previous works \cite{rodionova2021time,rodionovatcs22}, we analyze various properties of left (right) time robustness and propose an MILP encoding to control linear systems such that the left (right) time robustness is maximized. We continue along these lines and propose a novel notion of temporal robustness to account for both forward and backward temporal perturbations.

Besides the aforementioned notion of left (right) time robustness, there exist various other time robustness notions. Averaged STL was presented in  \cite{akazaki2015time} and captures temporal robustness by averaging spatial robustness over time intervals. Hybrid system conformance, see e.g., \cite{deshmukh2015quantifying,abbas2014formal}, quantifies the closeness of hybrid systems trajectories and measures a combination of spatial and time robustness, but does not allow for asynchronous time shift in the predicates.
\edit{The authors in \cite{buyukkocak2022temporal} 
	introduce a metric that can quantify
	the temporal relaxation of STL specifications.} 
Tailored to multi-agent systems, the authors  in \cite{sahin2017synchronous,sahin2019multirobot} propose counting linear temporal logic which requires a minimum number of agents for the satisfaction of a specification. The authors design control laws for such specifications where agents can implement their plans asynchronously, which can even account for time scaling effects, e.g., an agent pauses or speeds up, and not only time shifts in the predicate signal as we consider in this work. Temporal robustness of stochastic signals has been considered in \cite{lindemann2022temporal} by using risk measures, but the authors there consider time shifts in the system signal, opposed to time shifts in predicates. In \cite{selvaratnam2022mitl}, monitoring of  STL specifications under timing uncertainty in the underlying signal is considered by using over- and under-approximation of the satisfaction times of predicates. While \cite{baras_runtime} considers the time sensitive control for a subset of STL specifications, the authors in \cite{vasile2017time} present the time window temporal logic that is used in \cite{kamale2021automata,penedo2020language} to obtain control laws for finding temporal relaxations when the specification is not satisfiable. In \cite{chen2022stl}, the STL-based resiliency for cyber-physical system is presented that can capture temporal violations by recoverability and durability.

We make the following contributions. First, we propose a novel notion of temporal robustness  for STL specifications to account for forward and backward temporal perturbations. We  quantify the amount of permissible time shifts in the STL predicates to the left and right. We then show a set of desirable properties of our definition.  Furthermore, we propose an MILP encoding for control  of linear systems under the temporal robustness objective. 


\section{Signal Temporal Logic (STL)}
\label{sec:prelim}


\begin{figure}[t]
	\vspace{10pt}
	\centering
	\includegraphics[width=.48\textwidth]{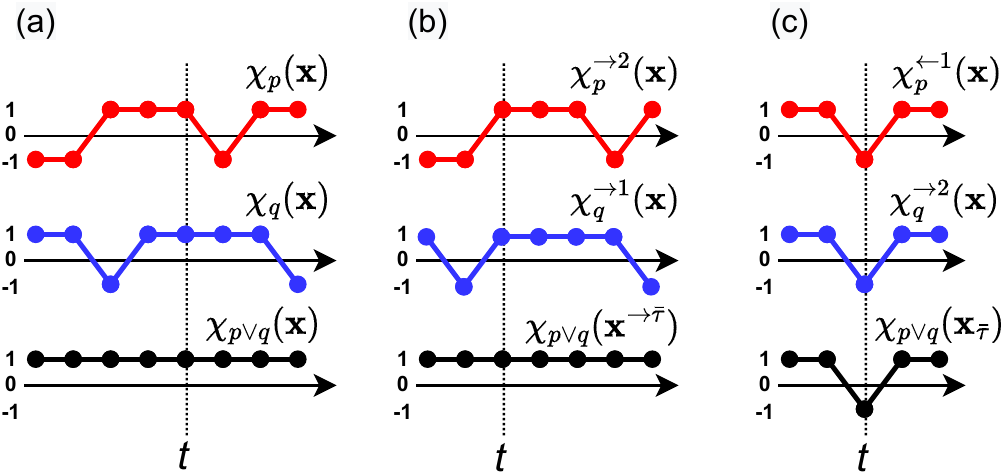}
	\caption{\small Predicates $p$, $q$ and STL formula $\varphi=p\vee q$ satisfaction over 
		(a) signal $\sstraj$; 
		(b) $\bar{\tau}$-late signal $\sstraj^{\rightarrow \bar{\tau}}$, where $\bar{\tau}=(2, 1)$; 
		(c) shifted signal $\sstraj_{\bar{\tau}}$, where $\bar{\tau}=(1, -2)$.  
		One can see that for signal $\sstraj$ and time $t$, the left and right time robustness are $\thetap_\varphi(\sstraj, t)=\thetam_\varphi(\sstraj,t) =2$. 
	}
	\label{fig:ex1}
\end{figure}


Let $\sstraj:\TDom \to X$ be a discrete-time signal with $\TDom \subseteq \Ne$ (we assume that $\Ne$ includes $0$) being the time domain and $x_t \in X$ being the state at time $t$, where $X\subseteq \Re^n$ is a metric space.
We call the set of all signals $\sstraj:\TDom \to X$ the \textit{signal space} $X^\TDom$.  A predicate $p$ is defined as $p\defeq \mu(x)\geq 0$, where $\mu(x): X\to\Re$ is a real-valued function of the state $x$. 
Let $I\subseteq\TDom$ be a time interval.
For any time point $t\in\TDom$, we define the set $t+I \defeq \{t+\tau\ |\, \tau\in I\}$. The syntax of Signal Temporal Logic (STL) is defined recursively as follows~\cite{maler2004monitoring}:
\begin{equation}
	\formula :\defeq p\ |\ \neg \formula \ |\ \formula_1 \land \formula_2 \ |\ \formula_1 \until_I \formula_2
\end{equation}
where 
$p\in AP$ is a predicate from a set of predicates $AP$, $\neg$ and $\land$ are the Boolean negation and conjunction, respectively, and $\until_I$ is the Until temporal operator over a time interval $I$. 
One can further define additional STL operators such as $\varphi_1\vee\varphi_2\defeq \neg(\neg\varphi_1\wedge\neg\varphi_2)$ (disjunction), $\eventually_I \varphi\defeq \top\until_I \varphi$ (eventually) and $\always_I \varphi \defeq \neg \eventually_I\neg \varphi$ (always). 

The  semantics of an STL formula $\varphi$ define when a signal $\sstraj$ satisfies $\varphi$ at time  $t$. Commonly, it is given via the STL characteristic function 
$\chi_\varphi(\sstraj,t): \SigSpace \times \TDom \to \{\pm 1\}$, see \cite{Donze10STLRob} for details. Intuitively, when $\chi_\varphi(\sstraj,t)=1$, it holds that the signal $\sstraj$ satisfies the formula $\varphi$ at time $t$, while $\chi_\varphi(\sstraj,t)=-1$
indicates that $\sstraj$ does not satisfy $\varphi$ at time $t$.


While the semantics of STL indicate \textit{if} the signal satisfies a given specification at time $t$, the \textit{left} and \textit{right} \textit{time robustness}   measures \textit{how robustly} a signal satisfies a given specification at time $t$ with respect to perturbations in time \cite{Donze10STLRob}.   
The left and right time robustness $\theta^\pm_\varphi(\sstraj,t)$ of a formula $\varphi$ relative to
a signal $\sstraj$ at time $t$  is defined recursively. For instance, the  left and right
 time robustness of a predicate $p$ are defined as follows:
	\begin{align*}
		\resizebox{\hsize}{!}{$\thetap_p(\sstraj,t) = \chi_p(\sstraj, t)\cdot
			\sup\{\tau\geq0:\forall t'\in[t,t\!+\!\tau],\chi_p(\sstraj,t')=\chi_p(\sstraj,t)\}$}
	\\
	\resizebox{\hsize}{!}{$\thetam_p(\sstraj,t) = \chi_p(\sstraj, t)\cdot
		\sup\{\tau\geq0:\forall t'\in[t\!-\!\tau,t],\chi_p(\sstraj,t')=\chi_p(\sstraj,t)\}$}
\end{align*}
and then, to obtain the $\theta^\pm_\varphi(\sstraj,t)$, one needs to apply the standard recursive $\inf$/$\sup$ rules to each $\theta^\pm_p(\sstraj,t)$, similarly to the characteristic function $\chi_\varphi(\sstraj,t)$, see \cite{rodionovatcs22} for details.


The sign of the left (right) time robustness reflects the satisfaction of the specification. Formally, if $\theta^\pm_\varphi(\sstraj, t)>0$ then $\chi_\varphi(\sstraj, t)=1$ and if $\theta^\pm_\varphi(\sstraj, t)<0$ then $\chi_\varphi(\sstraj, t)=-1$. 
In \cite{rodionovatcs22}, we also showed that the absolute value of the left (right) time robustness measures how robustly a signal $\sstraj$ satisfies a formula $\varphi$ at time $t$ with respect to time shifts in the predicates of formula $\varphi$. In fact, one can asynchronously shift predicates in time to the \textit{left} by up to $|\thetap_\varphi(\sstraj,t)|$ and the specification will not change its satisfaction. Formally, for $\tau_1,\ldots,\tau_K\in\mathbb{N}$, where $K$ is the number of predicates, if $\max_k\tau_k \leq |\thetap_\varphi(\sstraj,t)|$ then $\chi_\varphi(\sstraj^{\leftarrow\bar{\tau}}, t)=\chi_\varphi(\sstraj, t)$, where $\bar{\tau}=(\tau_1,\ldots,\tau_K)$ and $\sstraj^{\leftarrow\bar{\tau}}$ is a $\bar{\tau}$-early signal\footnote{The signal $\sstraj^{\leftarrow\bar{\tau}}$ is called a $\bar{\tau}$-early signal if $\forall p_k\in AP$, $\forall t\in\TDom$, $\chi_{p_k}(\sstraj^{\leftarrow\bar{\tau}}, t)=\chi_{p_k}(\sstraj, t+\tau_k)$. The signal $\sstraj^{\rightarrow\bar{\tau}}$ is called a $\bar{\tau}$-late signal if $\forall p_k\in AP$, $\forall t\in\TDom$, $\chi_{p_k}(\sstraj^{\rightarrow\bar{\tau}}, t)=\chi_{p_k}(\sstraj, t-\tau_k)$, see \cite{rodionovatcs22}.}. Analogously, if one shifts predicates in time to the \textit{right} by up to $|\thetam_\varphi(\sstraj, t)|$ then $\varphi$ will not change its satisfaction. Formally, for
$\tau_1,\ldots,\tau_K\in\Ne$, if $\max_k\tau_k \leq |\thetam_\varphi(\sstraj,t)|$ then $\chi_\varphi(\sstraj^{\rightarrow\bar{\tau}}, t)=\chi_\varphi(\sstraj, t)$, where $\sstraj^{\rightarrow\bar{\tau}}$ is a $\bar{\tau}$-late signal.


\edit{
\begin{exmp}\label{ex:1}
	In Fig.~\ref{fig:ex1}(a), we  plotted a characteristic function of two predicates $p$ and $q$ and the formula $\varphi:=p\vee q$. 
	The right time robustness is  $\thetam_{p\vee q}(\sstraj, t) = 2$ (since $\thetam_p(\sstraj, t) = 2$ and $\thetam_q(\sstraj, t)=1$). 
	Hence,  the predicates can be shifted by up to $2$ time steps  to the \textit{right} and the formula at time  $t$ must still be satisfied, see Fig.~\ref{fig:ex1}(b).
	The left time robustness is $\thetap_{p\vee q}(\sstraj, t) = 2$ (since $\thetap_p(\sstraj, t) = 0$ and $\thetap_q(\sstraj, t)=2$). The predicates can thus be shifted by up to $2$ time steps to the \textit{left} and the formula must still be satisfied.
\end{exmp}
}
\section{Temporal Robustness}
\label{sec:comb_robustness}

Note that the left (right) time robustness is directional: its value provides a bound on how much  predicates can be shifted to the left (right). Importantly, one cannot  consider time shifts of some predicates to the left, while some other predicates are shifted to the right.
\edit{For instance, note that if we shift a predicate $p$ in Fig.1(a) by 1 time step to the \textit{left}, but a predicate $q$ by 2 time steps to the \textit{right}, see Fig.~\ref{fig:ex1}(c), then for the shifted signal $\sstraj_{\bar{\tau}}$, where $\bar{\tau}=(1, -2)$,
	the formula satisfaction at time $t$ changes, i.e., it holds that
	$\chi_\varphi(\sstraj_{\bar{\tau}}, t)=-1\not = \chi_\varphi(\sstraj, t)$. }
To overcome this limitation, we propose \edit{a} temporal robustness  which quantifies the amount of permissible time perturbation in both directions.  

\begin{definition}
	\label{def:temporal_rob}
	The \emph{temporal robustness} $\theta_\varphi(\sstraj,t)$ of an STL formula $\varphi$ relative to a signal $\sstraj:\TDom \rightarrow X$ at time $t \in \TDom$ is defined recursively as follows:
	\begin{equation*}
	\begin{aligned}
		\theta_p(\sstraj,t) &\defeq \chi_p(\sstraj, t)\cdot
		\sup\{\tau\geq0\ :\ \forall t' \text{ s.t. } |t'-t|\leq \tau,\\ &\qquad\qquad\qquad\qquad\qquad\quad\chi_p(\sstraj,t')=\chi_p(\sstraj,t)\}\\
	\theta_{\neg \formula}(\sstraj,t) &\defeq -\theta_\formula(\sstraj,t)
	\\
	\theta_{\formula_1 \land \formula_2}(\sstraj,t) &\defeq \inf\left(\theta_{\formula_1}(\sstraj,t),\ \theta_{\formula_2}(\sstraj,t)\right)
	\\
	\theta_{\formula_1 \until_I \formula_2}(\sstraj,t) &\defeq 
	\sup_{t'\in t+I} \inf\left(
	\theta_{\formula_2}(\sstraj,t'),\
	\inf_{t'' \in [t,t')} \theta_{\formula_1}(\sstraj,t'')\right)
	\end{aligned}
	\end{equation*}
When robustness is evaluated at $t=0$, we denote it as $\theta_\varphi(\sstraj)$ as a shorthand notation for $\theta_\varphi(\sstraj, 0)$. 
\end{definition}

 We next show soundness of our definition, and remark that the proofs of our results   are provided in the Appendix.

\begin{theorem}[Soundness]
	\label{thm:sat_theta}
	For an STL formula $\varphi$, signal $\sstraj:\TDom \rightarrow X$ and some time $t\in\TDom$, it holds that
	\begin{enumerate}
		\item\label{i1} If $\theta_\varphi(\sstraj,t) > 0$, then $\chi_\varphi(\sstraj,t)= +1$. 
		\item\label{i2}  If
		$\theta_\varphi(\sstraj,t) < 0$, then $\chi_\varphi(\sstraj,t)= -1$.
	\end{enumerate}
\end{theorem}


Let us next analyze what information $\theta_\varphi(\sstraj, t)$ gives us about  robustness. First going back to Example \ref{ex:1} and Fig.~\ref{fig:ex1}(a), the temporal robustness is $\theta_{p\vee q}(\sstraj, t) = 1$ (since $\theta_p(\sstraj, t) = 0$ and $\theta_q(\sstraj, t)=1$) which gives us  the  desired result that the left (right) time robustness could not give us.  Recall that we  consider temporal robustness by time shifts in the characteristic functions $\chi_{p_k}(\sstraj, t)$ in an asynchronous manner, i.e., for each predicate $p_k$ individually. Formally, 
for time shifts $\tau_1,\ldots,\tau_K\in\Ze$,
we say that a signal $\sstraj_{\bar{\tau}}$ is an 
\emph{asynchronously shifted signal} 
if $\chi_{p_k}^{}(\sstraj_{\bar{\tau}}, t)=
	\chi_{p_k}(\sstraj, t+\tau_k)$
for all $t\in\TDom$ and for all $ p_k\in AP=\{p_1,\ldots,p_K\}$. We next show how the 
temporal robustness $\theta_\varphi(\sstraj, t)$ relates to permissible time shifts $\tau_k$ via $\sstraj_{\bar{\tau}}$.


\begin{theorem}
	\label{thm:shift_async}
	Let $\varphi$ be an STL formula built upon a predicate set $AP=\{p_1,\ldots,p_K\}$. Let $\sstraj:\TDom \rightarrow X$ be a signal and $t\in\TDom$ be a time point. For $\forall\tau_1,\ldots,\tau_K\in\Te$, it holds that:
	\begin{equation*}
		\max(|\tau_1|,\ldots,|\tau_K|) \leq |\theta_\varphi(\sstraj,t)| \ \ \Longrightarrow\ \  
		\chi_\varphi(\sstraj_{\bar{\tau}},t )=\chi_\varphi(\sstraj,t).	
	\end{equation*}
\end{theorem}

For predicates, we show an interesting connection between the temporal robustness and the left (right) time robustness which follows directly from the definition.

\begin{cor}
	\label{cor:theta_pm}
	Given a predicate $p\in AP$ and a signal $\sstraj:\TDom \rightarrow X$, for any $t\in\TDom$, the following equality holds:
	\begin{equation}
		\label{eq:theta_connection}
		\theta_p(\sstraj,t) =
		\chi_p(\sstraj,t)\cdot\min\left(|\thetap_p(\sstraj,t)|,\ |\thetam_p(\sstraj,t)|\right)
	\end{equation}
\end{cor}
For a formula $\varphi$, it however does not hold that $\theta_\varphi(\sstraj,t) =
\chi_\varphi(\sstraj,t)\cdot\min\left(|\thetap_\varphi(\sstraj,t)|,\ |\thetam_\varphi(\sstraj,t)|\right)$, e.g., as in Example \ref{ex:1}. However, we can prove the following relation between them.


%

\begin{theorem}
	\label{thm:bound}
	Given an STL formula $\varphi$ and a signal $\sstraj:\TDom \rightarrow X$, for any $t\in\TDom$,
	$|\theta_\varphi(\sstraj,t)| \leq
	|\theta^\pm_\varphi(\sstraj,t)|$.	
\end{theorem}


\begin{figure*}[t!]
	\vspace{10pt}
	\setkeys{Gin}{height=57mm}
	\subfloat[$J=\theta_\varphi(\sstraj)$: Temporal robustness maximization. 	
	Found maximum is $\theta_\varphi(\sstraj^*)=4$. Evaluated left and right time robustness values are $\thetap_\varphi(\sstraj^*)=6$ and $\thetam_\varphi(\sstraj^*)=4$ respectively.]{\includegraphics{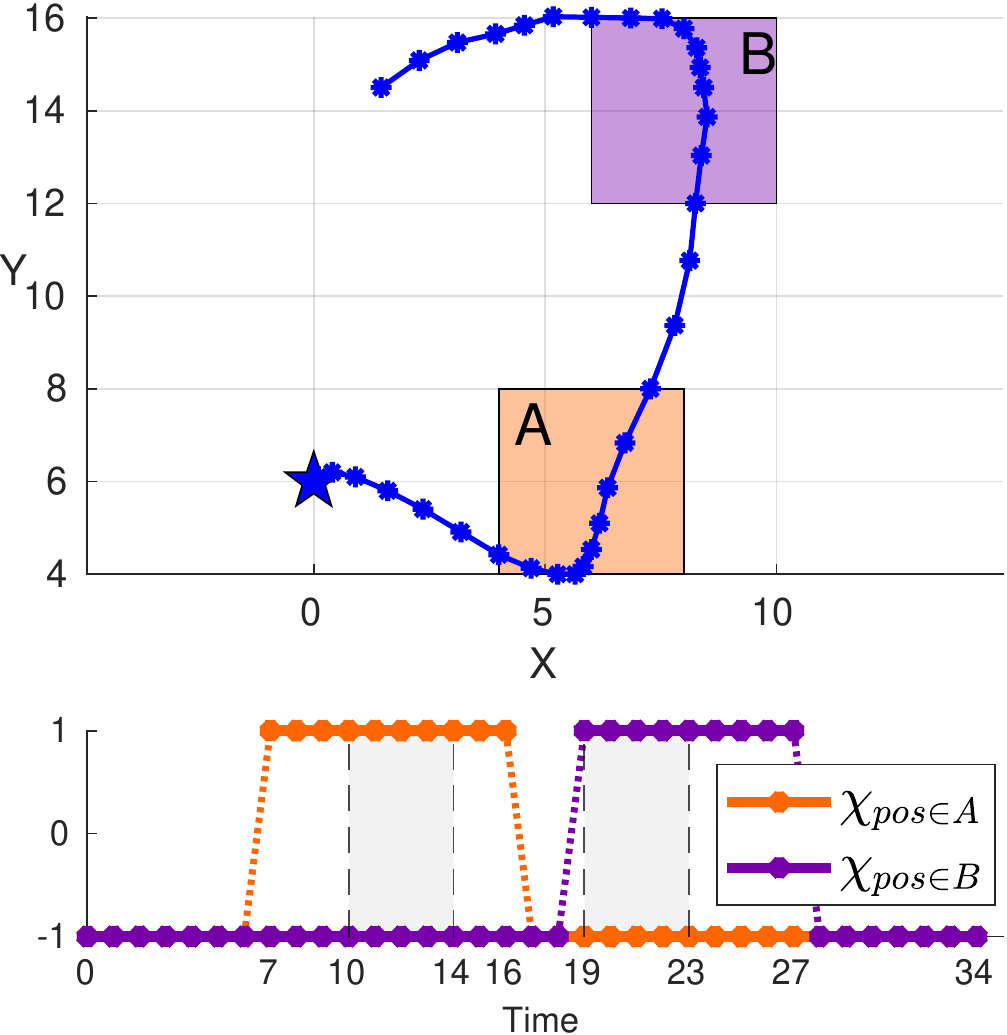}}
	\hfill
	\subfloat[$J=\thetap_\varphi(\sstraj)$: Left time robustness maximization. 	
	Found maximum is $\thetap_\varphi(\sstraj^*)=10$. Evaluated temporal robustness and right time robustness values are $\theta_\varphi(\sstraj^*)=\thetam_\varphi(\sstraj^*)=0$.]{\includegraphics{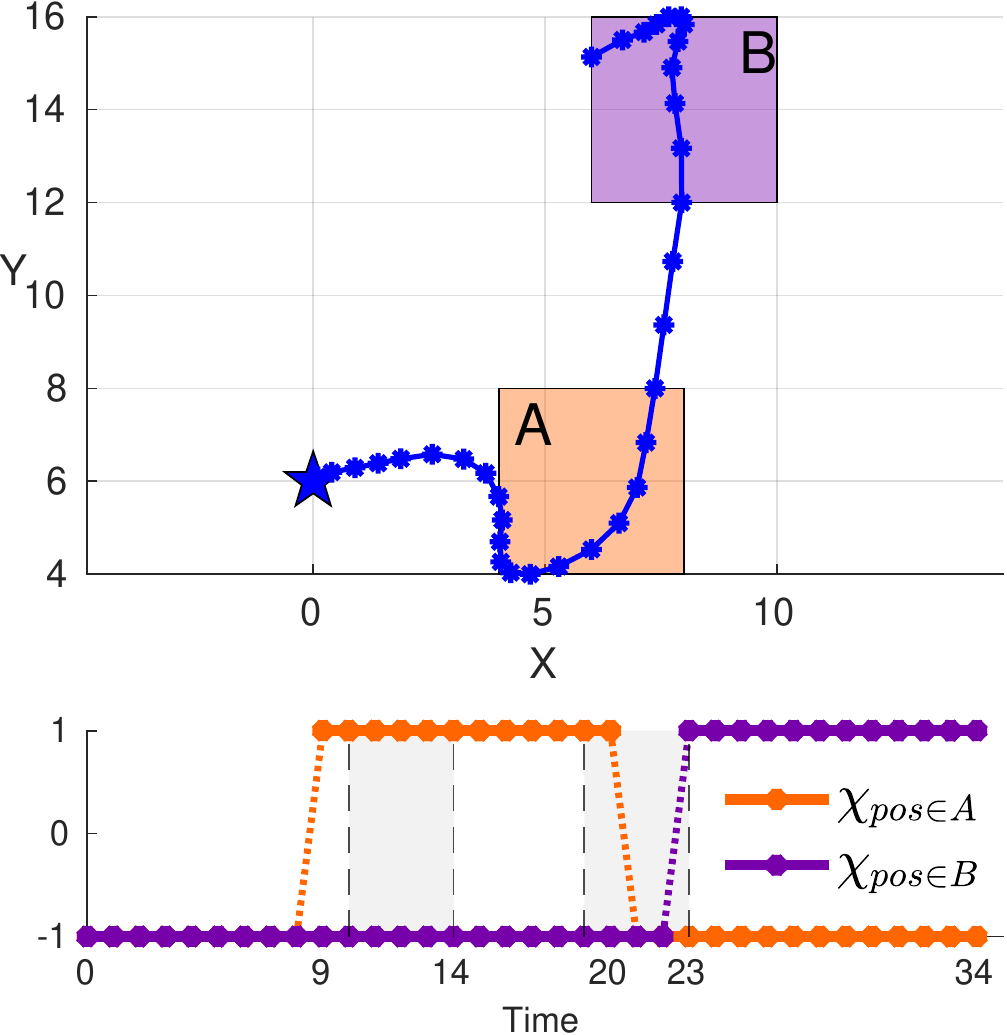}}
	\hfill
	\subfloat[$J=\thetam_\varphi(\sstraj)$: Right time robustness maximization. 	
	Found maximum is $\thetam_\varphi(\sstraj^*)=6$. Evaluated temporal robustness and left time robustness values are $\theta_\varphi(\sstraj^*)=\thetap_\varphi(\sstraj^*)=3$.]{\includegraphics{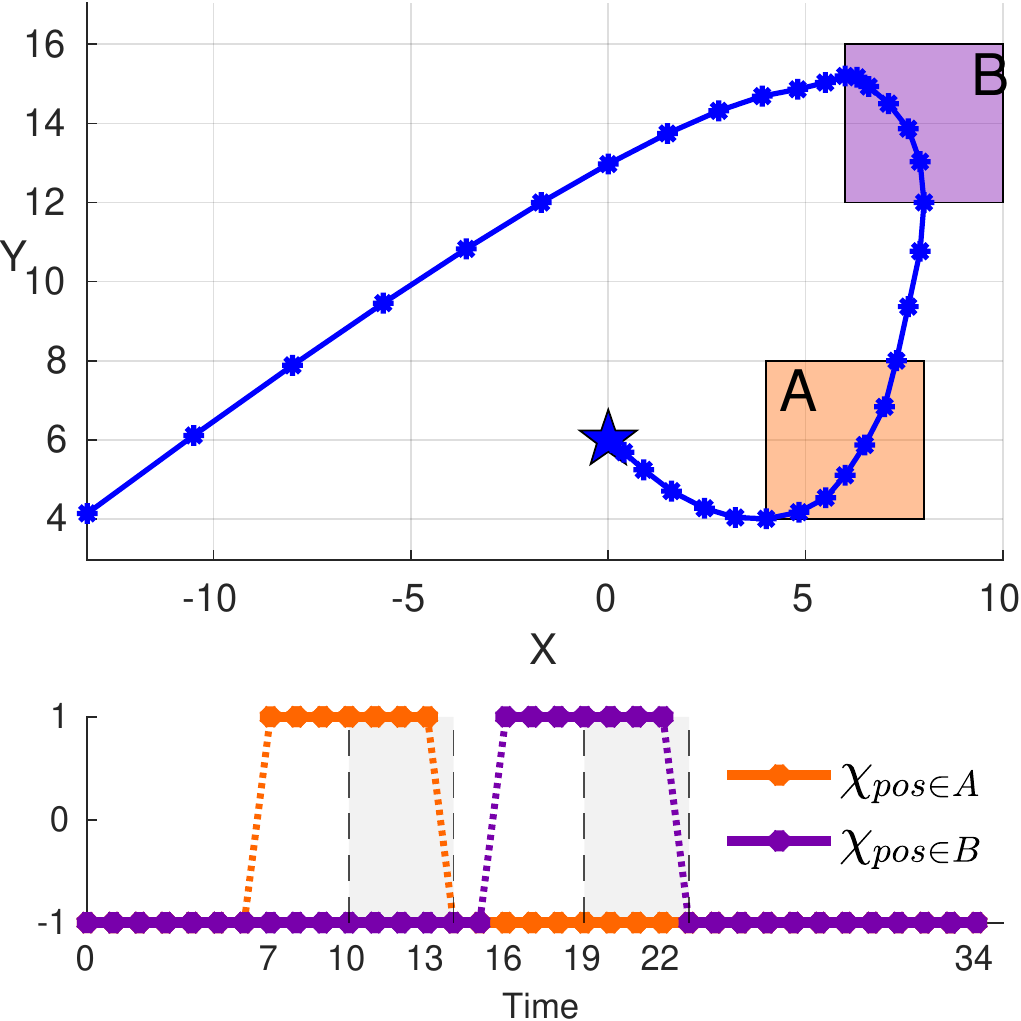}}
	\caption{Timed Navigation. Maximization of various temporal robustness objectives $J$.}
	\label{fig:exp1}
	\vspace{-4mm}
\end{figure*}

\section{Temporally-Robust STL Control Synthesis}
\label{sec:synthesis}
 
 Let us next address the question of how to control a system to be temporally robust.  We particularly consider linear systems and assume that the formula $\varphi$ is build upon linear predicates. Our goal is to find an optimal control sequence $\inpSig^*$ such that the corresponding trajectory $\sstraj$ respects  input and state constraints and satisfies the specification $\varphi$ robustly while maximizing a desired cost function $J$.
 

\begin{prob}[STL Control Synthesis]
	\label{prob:synthesis}
	Given an STL specification $\varphi$, time horizon\footnote{We assume that $\varphi$ are bounded-time STL formulas with formula length $len(\varphi) \leq H$. For the formula length definition, we refer the reader to \cite{raman2014model}.} $H$, discrete-time linear control system with initial condition $x_0\in X_0$, 
	solve
	\begin{equation*}
		\label{eq:control_problem}
		\begin{aligned}
			\inpSig^*=\underset{\inpSig}{\argmax}
			&\quad J\left(x_0,\,\inpSig,\,\sstraj,\,\varphi\right)
			\\
			\text{s.t.}
			&\quad x_{t+1} = \dynA x_t + \dynB u_t,\ u_t\in U,\ t=0,\ldots,H-1\\
			&\quad x_0\in X_0,\quad x_t\in X,\ t=1,\ldots,H\\
			&\quad \chi_\varphi(\sstraj)=1
		\end{aligned}
	\end{equation*}
	where $J\left(x_0,\,\inpSig,\,\sstraj,\,\varphi\right)$ is the desired cost function. 
	In \textit{robust STL control synthesis} the cost function depends on a specific robustness of interest, e.g. spatial robustness~\cite{fainekos2009robustness}, left (right) time robustness~\cite{rodionova2021time}, and in our particular case, temporal robustness.
\end{prob}	

To solve Problem \ref{prob:synthesis} with $J = \theta_\varphi(\sstraj)$, we present a mixed-integer linear (MILP) encoding of the temporal robustness $\theta_\varphi(\sstraj)$. Recall that by Def.~\ref{def:temporal_rob}, $\theta_\varphi(\sstraj,t)$ is defined
 recursively on the structure of $\varphi$. 
 Below, we describe the main milestone of the overall MILP encoding, that is the
 encoding of predicates $\theta_p(\sstraj, t)$. From Cor.~\ref{cor:theta_pm} and Thm.~\ref{thm:sat_theta}, we get that
\begin{equation}
	\label{eq:cases}
	\theta_p(\sstraj,t) = 
	\begin{cases}
	\min\left(\thetap_p(\sstraj,t),\, \thetam_p(\sstraj,t)\right) &\text{if }\chi_p(\sstraj, t)=1\\
	\max\left(\thetap_p(\sstraj,t),\, \thetam_p(\sstraj,t)\right) &\text{if }\chi_p(\sstraj, t)=-1
	\end{cases}
\end{equation}
\edit{The complete MILP encoding of $\thetap_p(\sstraj,t)$, $\thetam_p(\sstraj,t)$ and $\chi_p(\sstraj, t)$ is presented in \cite{rodionova2021time}. The encoding in \cite{rodionova2021time} introduces binary variables $z_t\in\{0,1\}$ to represent the Boolean satisfaction of the given predicate $p$ at every time point $t$ within the horizon and also introduces the integer counter variables $c_t^1$ and $c_t^0$ to enumerate sequential time points in the future and in the past for which $\chi_p(\sstraj, t)$ does not change its value.}

\edit{Next, having encoded the left $\thetap_p(\sstraj,t)$ and right $\thetam_p(\sstraj,t)$ temporal robustness of a predicate $p$, the $\min$ and $\max$ operators used in \eqref{eq:cases} can be encoded utilizing the rules from \cite{raman2014model}.
	For instance, if $\thetap_p(\sstraj, t)=r_1$ and $\thetam_p(\sstraj, t)=r_2$, then $\min(\thetap_p(\sstraj, t),\ \thetam_p(\sstraj, t))=r$ if and only if:
\begin{equation}
\begin{aligned}
	&r_i-M(1-b_i) \leq r \leq r_i,\quad \forall i\in\{1,2\}\\
	&b_1+b_2=1
\end{aligned}
\vspace{-3pt}
\end{equation}
where $b_i=\{0,1\}$ are introduced binary variables and $M$ is a big-$M$ parameter. The $\max$ operator can be encoded similarly.
}

\edit{Thus,
we obtain the MILP encoding of the two variables from \eqref{eq:cases}, 
$\edit{\mintheta_t}\defeq	\min\left(\thetap_p(\sstraj,t),\ \thetam_p(\sstraj,t)\right)$
and 
$\edit{\maxtheta_t}\defeq	\max\left(\thetap_p(\sstraj,t),\, \thetam_p(\sstraj,t)\right)$.}
\edit{
Using \eqref{eq:cases} and the binary variables $z_t\defeq \frac{\chi_p(\sstraj, t)+1}{2}$, the temporal robustness $\theta_p(\sstraj, t)$ is defined as\footnote{Note that \eqref{eq:5} can be expressed
	as a set of MILP constraints according to \cite[Lemma 4.1]{rodionova2021time}.}
\begin{equation}
	\label{eq:5}
	\theta_p(\sstraj, t) = z_t\mintheta_t  + (1-z_t)\maxtheta_t.
	\vspace{-3pt}
\end{equation}
}
We can now use the MILP encoding for the remaining Boolean and temporal operators as originally presented in \cite{raman2014model}. \edit{In Section \ref{sec:experiments} and Table~\ref{tab:my-table} we present a comparison analysis of the performance and computation times of solving Problem~\ref{prob:synthesis} for various temporal robustness functions, such as $J = \theta_\varphi(\sstraj)$ and $J = \theta^\pm_\varphi(\sstraj)$}.	

\section{Experimental Results}
\label{sec:experiments}

\begin{table}[]
	\vspace{8pt}
	\renewcommand{\arraystretch}{1.5}
	\setlength{\tabcolsep}{3.pt}
	\centering
	\begin{tabular}{|c|l|c|l|}
		\hline
		\textbf{Mission}               & \textbf{Objective} $J$ & \textbf{\edit{Comp. Time}} (s) & \textbf{Simulations} \\ \hline\hline
		\multirow{3}{*}{Scen. 1}& $\theta_\varphi(\sstraj^*)=4$           & $12.36$         & \href{https://tinyurl.com/temp-rob}{https://tinyurl.com/temp-rob}         \\ \cline{2-4} 
			& $\thetap_\varphi(\sstraj^*)=10$        & $2.95$          & \href{https://tinyurl.com/temp-left}{https://tinyurl.com/temp-left}         \\ \cline{2-4} 
			& $\thetam_\varphi(\sstraj^*)=6$           & $4.78$        & \href{https://tinyurl.com/temp-right}{https://tinyurl.com/temp-right}         \\ \hline\hline
			Scen. 2              & $\theta_\varphi(\sstraj^*)=3$           & $35.74$          & \href{https://tinyurl.com/uav-surv}{https://tinyurl.com/uav-surv}         \\ \hline
		\end{tabular}
		\caption{\small Summary of experimental results.}
		\label{tab:my-table}
		\vspace{-6mm}
	\end{table}

In this section, we present two case studies in which we solve the control-synthesis problem \ref{prob:synthesis} for various cost functions. All simulations were performed on an Intel Core i7-9750H 6-core processor with 16GB RAM. The code was implemented in MATLAB using \edit{YALMIP}~\cite{lofberg2004yalmip} with Gurobi 9.1~\cite{gurobi} as
the solver. \edit{The computation times and links to animations are reported in Table~\ref{tab:my-table}}.

\textbf{Scenario 1 - Timed Navigation.} Consider an autonomous agent with 2D position and velocity $x\defeq (\pos,\, \vel)\in\Re^4$ where $(\pos_0,\vel_0) \defeq (0, 6,0,0)$. We consider the dynamics
\begin{equation}
	\label{eq:sys}
	x_{t+1}=\dynA x_t+\dynB u_t, \quad ||u_t||_\infty \leq 20	
\end{equation}
where $\dynA\defeq I_{2} \otimes \begin{bmatrix}
	1 &0.1\\ 0&1
\end{bmatrix}$ and $\dynB\defeq\begin{bmatrix}
	0.005\\0.1 \end{bmatrix}$.  The agent should first reach zone $A$, see Fig.~\ref{fig:exp1} for an illustration, any time within the time interval $[10, 14]$ and then reach zone $B$ any time within $[19, 23]$ as captured by
the  STL specification:
\begin{equation}
	\varphi \defeq \eventually_{[10,14]} \left(\pos\in A\right) \ \wedge\  \eventually_{[19,23]} \left(\pos\in B\right)
\end{equation}
where  $ A\defeq (x\geq 4) \wedge (x\leq 8) \wedge (y\geq 4)\wedge (y\leq 8)$ 
and $ B\defeq (x\geq 6)  \wedge(x\leq 10) \wedge (y\geq 12)\wedge (y\leq 16)$.

We first solve Problem \ref{prob:synthesis} for $J=\theta_\varphi(\sstraj)$ and plot the resulting trajectory $\sstraj^*$ in Fig.~\ref{fig:exp1}(a), and obtain $\theta_\varphi(\sstraj^*)=4$. From the characteristic function plotted in Fig.~\ref{fig:exp1}(a), one can see that if the agent starts the execution of the trajectory by up to 4 time steps earlier or later, the mission specification $\varphi$ will still be satisfied, since for such a shifted trajectory there will be at least one point in time, where the agent is within zone $A$ and $B$ within the specified time intervals (depicted in grey color). This result \edit{supports} Thm.~\ref{thm:shift_async} derived previously. For comparison, the calculated left and right time robustness are $\thetap_\varphi(\sstraj^*)=6$ and $\thetam_\varphi(\sstraj^*)=4$, respectively. One can see, that indeed, 	$\theta_\varphi(\sstraj^*) \leq
\theta^\pm_\varphi(\sstraj^*)$ which is expected by Thm.~\ref{thm:bound}.

To compare the system's behavior under different cost functions in Problem~\ref{prob:synthesis}, we use the left time robustness   $J=\thetap_\varphi(\sstraj)$ and the right time robustness $J=\thetam_\varphi(\sstraj)$ as control objectives. We next show that the temporal robustness is preferred over the left and right time robustness when dealing with systems where the direction of perturbations in time is unknown.

The results of maximizing the left time robustness are presented in Fig.~\ref{fig:exp1}(b) where
$J^*=\thetap_\varphi(\sstraj^*)=10$. It is expected that the maximization of the left time robustness leads to a trajectory for which the agent reaches the desired goal within the required time bounds and then it stays there for as long as possible. In Fig.~\ref{fig:exp1}(b) this is represented as $\chi_{\pos\in A}(\sstraj^*, 9) = \ldots=\chi_{\pos\in A}(\sstraj^*, 20)=1$ and $\chi_{\pos\in B}(\sstraj^*, 23) = \ldots=\chi_{\pos\in B}(\sstraj^*, 34)=1$. This means that if the agent starts the execution earlier by up to 10 time units (the trajectory is shifted to the left), the mission will still be satisfied. However, any perturbation that leads to a  trajectory shifted to the right results in a violation of the specification, $\theta_{\varphi}(\sstraj^*)=\thetam(\sstraj^*)=0$. Indeed, in this case, the agent will not be able to visit the zone $B$ within $[19, 23]$ time units, see Fig.~\ref{fig:exp1}(b). 

The results of maximizing the right time robustness are presented in Fig.~\ref{fig:exp1}(c) where $J^*=\thetam_\varphi(\sstraj^*)=6$.
Note that in this case, the agent reaches both zones as soon as possible, see Fig.~\ref{fig:exp1}(c).  We obtain the temporal robustness and left time robustness of $\theta_\varphi(\sstraj^*)=\thetap_\varphi(\sstraj^*)=3$. We can again see that $\theta_\varphi(\sstraj^*)\leq \theta^\pm_\varphi(\sstraj^*)$ which \edit{is consistent with} Thm.~\ref{thm:bound}. 
Also note that since the evaluated left time robustness $\thetap_\varphi(\sstraj^*)=3$, only the predicate shifts up to $3$ time steps to the left still guarantee the satisfaction of the specification. From Fig.~\ref{fig:exp1}(c) one can see that the shift by $4$ time steps to the left leads to an agent leaving both regions of interest sooner than the predefined intervals, therefore, the mission is violated. 

\begin{figure*}[h!]
	\vspace{10pt}
	\setkeys{Gin}{height=36mm}
	\subfloat{\includegraphics{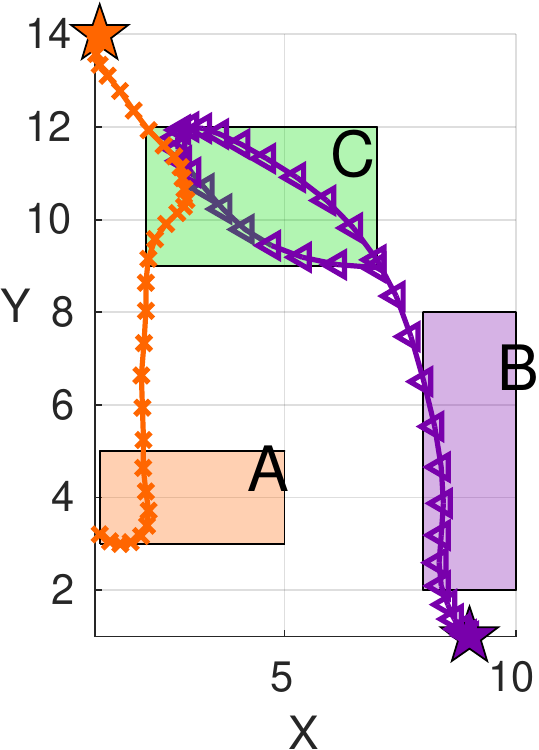}}
	\hfill
	\subfloat{%
		\includegraphics{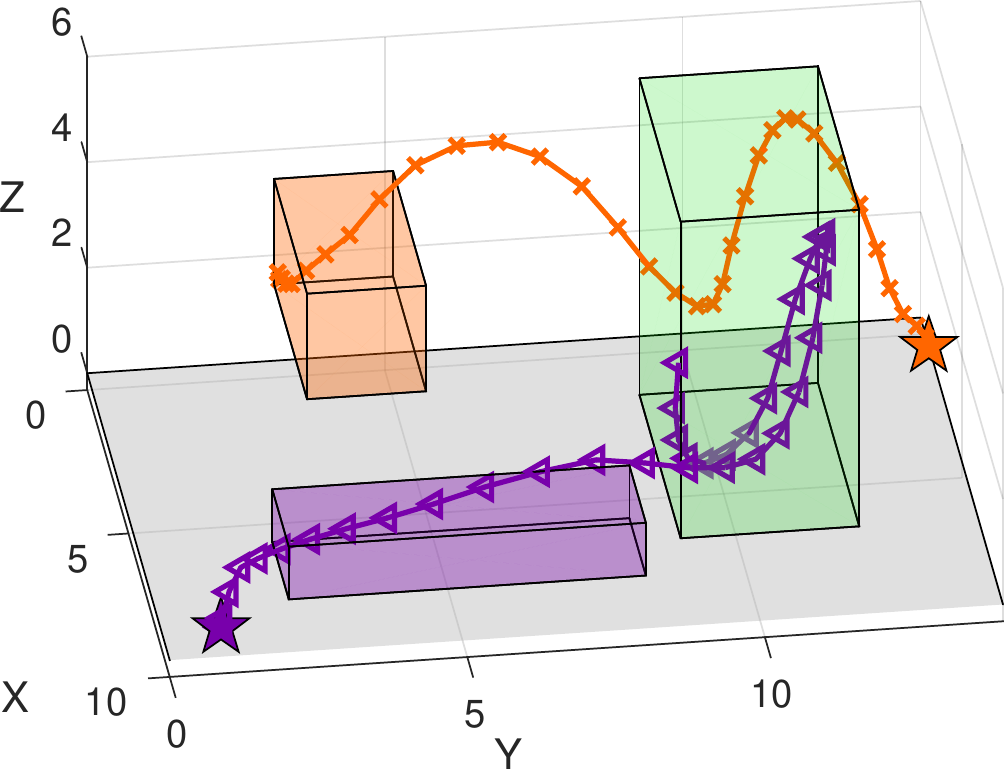}}
	\hfill
	\subfloat{%
		\includegraphics{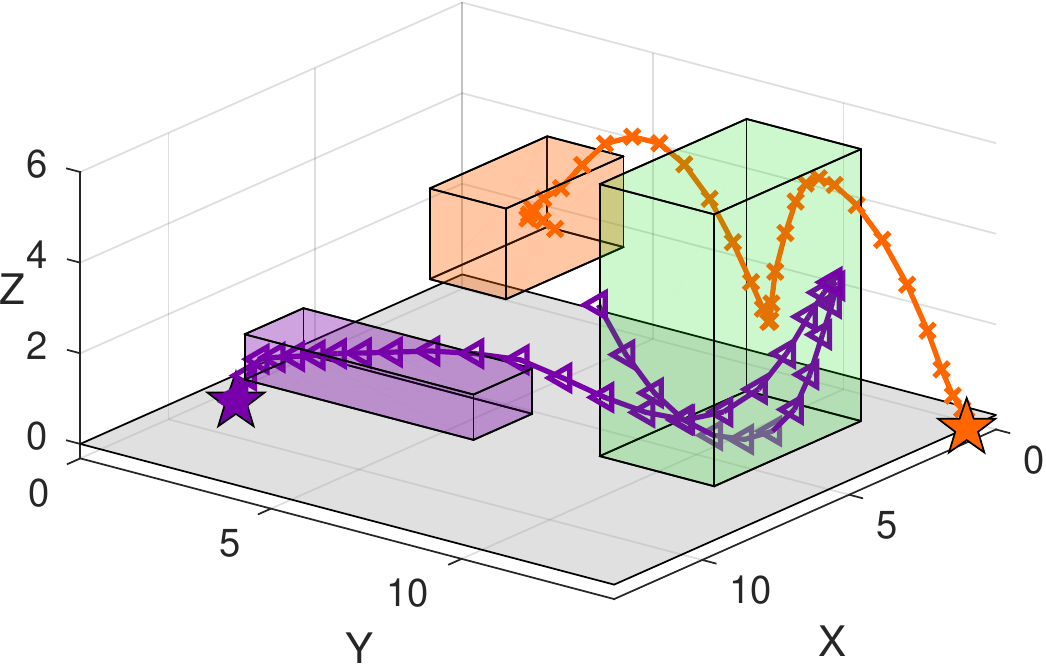}}
	\hfill
	\subfloat{%
		\includegraphics{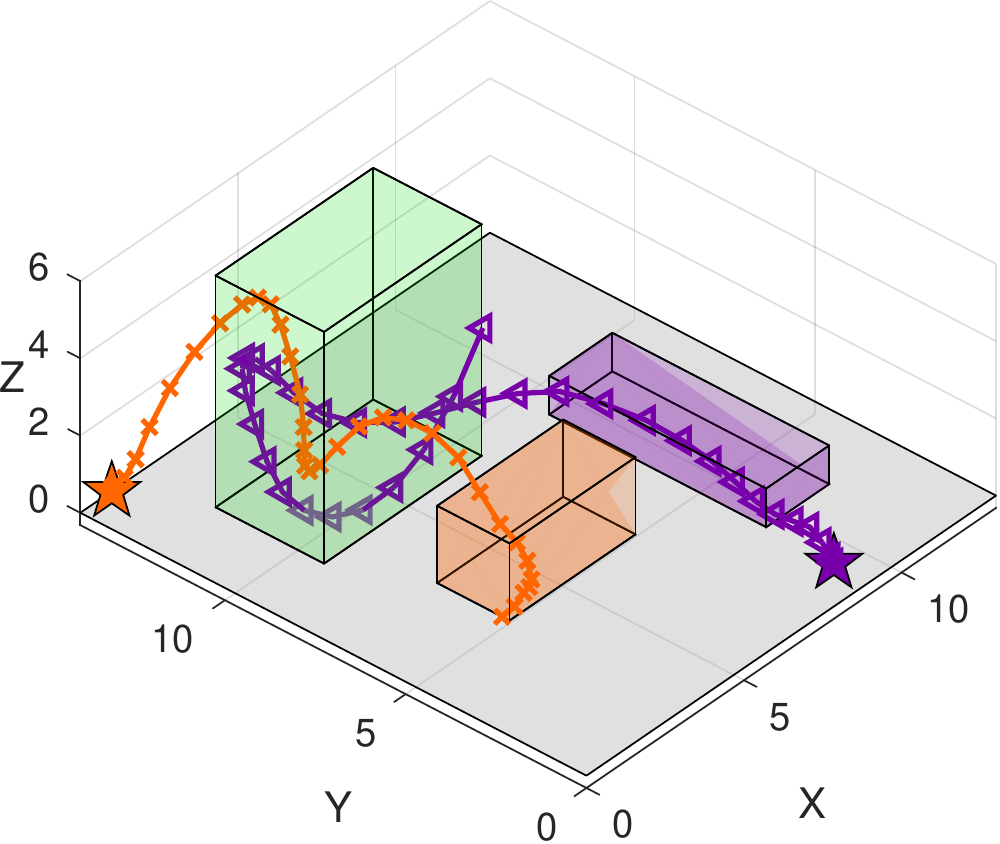}}
	\caption{\small Timed Multi-UAV Surveillance. 2D and 3D representation of the map from different angles. Optimal trajectory $\sstraj^*$ is obtained by the maximization of  temporal robustness using Problem~\ref{prob:synthesis}. 	
		Found maximum temporal robustness is $\theta_\varphi(\sstraj^*)=3$.}
	\label{fig:exp2_map}
	\vspace{-2mm}
\end{figure*}

\begin{figure}[h!]
	\centering
	\includegraphics[width=0.445\textwidth]{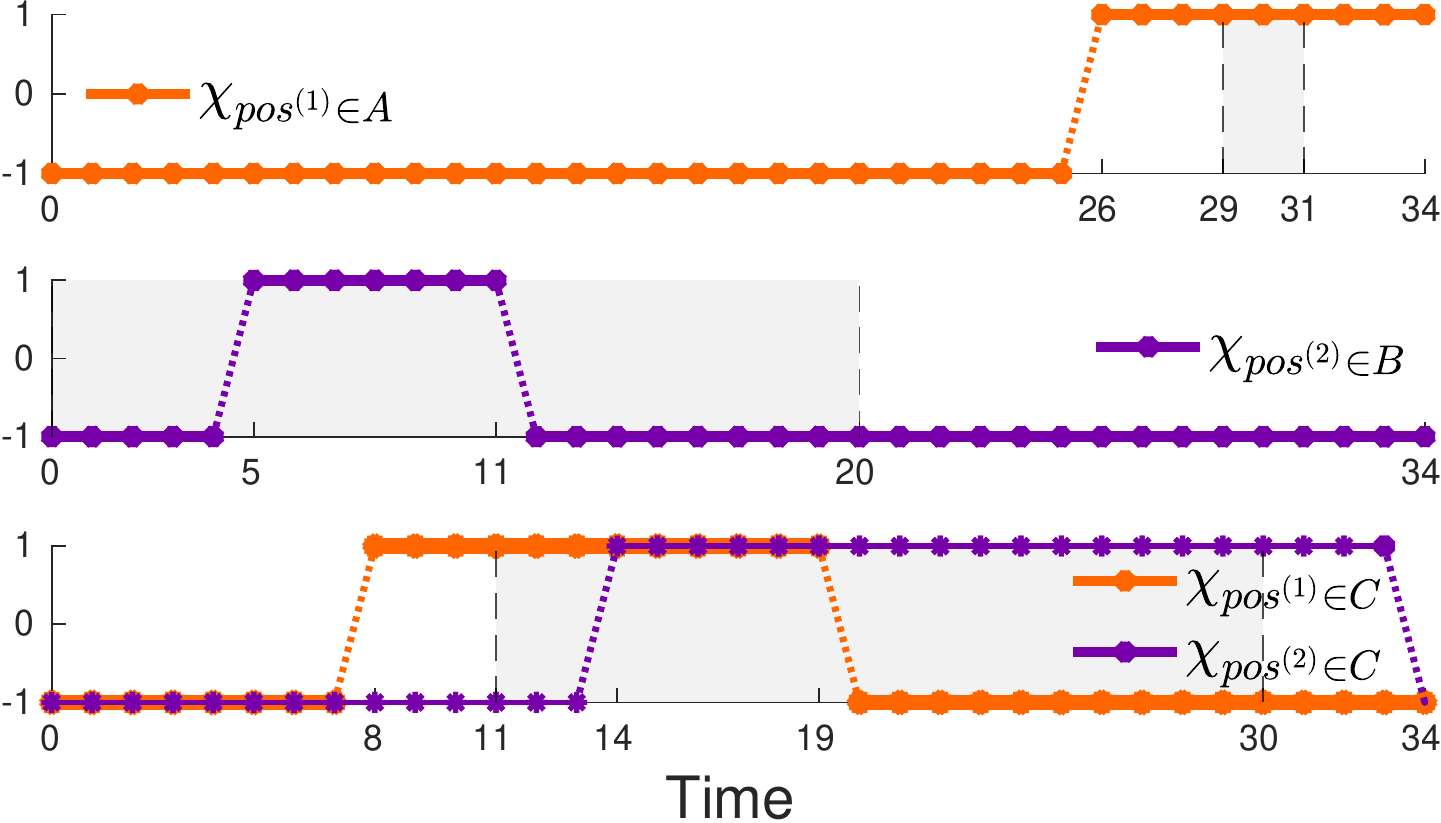}
	\caption{\small Timed Multi-UAV Surveillance. 
		Various sub-formulas satisfaction over an optimal signal $\sstraj^*$.
		Found maximum temporal robustness is $\theta_\varphi(\sstraj^*)=3$.}
	\label{fig:exp2_pred}
	\vspace{-3mm}
\end{figure}

\textbf{Scenario 2 - Timed Multi-UAV Surveillance.} We now consider two unmanned aerial vehicles (UAVs)
in a surveillance mission. Particularly, consider the $d$th agent with state $x^{(d)} \defeq (\pos^{(d)}, \vel^{(d)}) \in \Re^6$ where $\pos$ and $\vel$ are the 3D position and velocity, see Fig.~\ref{fig:exp2_map}. The initial states are set to be $(\pos_0^{(1)},\vel_0^{(1)})\defeq(1, 14,0, 0, 0,0)$ and $(\pos_0^{(2)},\vel_0^{(2)})\defeq(9, 1,0, 0, 0,0)$. Let the dynamics of both UAVs be of the form $x^{(d)}_{t+1} = \dynA x^{(d)}_t + \dynB u^{(d)}_t$ where  $\dynA$ and $\dynB$ are obtained through the linearization of
the UAV dynamics, see \cite{luukkonen2011modelling} for more details. The inputs $u^{(d)}_t \in \Re^3$ are the thrust, roll, and pitch of the UAV. 

The UAVs are tasked with a persistent surveillance mission of
the region $C$, see Fig.~\ref{fig:exp2_map}, while each of them must visit their individually assigned regions $A$ and $B$.  The overall specification is of the form $\varphi\defeq \bigwedge_{i=1}^3 \varphi_i$ where:

\begin{enumerate}
	\item UAV 1 should reach and stay in zone $A$ all the time from $29$ to $31$ time units, $\varphi_1 \defeq \always_{[29, 31]} \left(\pos^{(1)}\in A\right)$.
	\item UAV 2 should eventually reach zone $B$ any time between $0$ and $20$ time units, $\varphi_2\defeq\eventually_{[0,20]} \left(\pos^{(2)}\in B\right)$.
	\item Region $C$ should be surveilled, i.e. either one or both UAVs must be within $C$ all the time from $11$ to $30$ time units,
	 $\varphi_3\defeq\always_{[11, 30]} \left(\pos^{(1)}\in C\ \vee\ \pos^{(2)}\in C\right)$.	 
\end{enumerate}

Similarly to the 2D case, the regions $A$, $B$ and $C$ are defined via a set of conjunctions over linear predicates.

We solve Problem~\ref{prob:synthesis} for the temporal robustness objective which leads to the optimal solution $J^*=\theta_{\varphi}(\sstraj^*)=3$. 
Such optimal solution due to Thm.~\ref{thm:shift_async} guarantees that for any shifted signal $\sstraj_{\bar{\tau}}$ with $\max(|\tau_1|, |\tau_2|) \leq 3$, the mission specification will be satisfied. Take a look at Fig.~\ref{fig:exp2_pred}. For the corner case, if one shifts the orange line to the left by $3$ time units and the violet one to the right by $3$ time units, i.e. $\bar{\tau}=(3, -3)$, then one can see that $\chi_{\pos^{(1)}\in C}(\sstraj_{\bar{\tau}}, 5) =\ldots = \chi_{\pos^{(1)}\in C}(\sstraj_{\bar{\tau}}, 16)=1$ and 
$\chi_{\pos^{(2)}\in C}(\sstraj_{\bar{\tau}}, 17) =\ldots = \chi_{\pos^{(2)}\in C}(\sstraj_{\bar{\tau}}, 34)=1$, therefore, $\chi_{\varphi_3}(\sstraj_{\bar{\tau}})=1$. Analogously, $\varphi_1$ and $\varphi_2$ are satisfied by $\sstraj_{\bar{\tau}}$, therefore, the overall satisfaction of $\varphi$ is indeed preserved by the shift $\bar{\tau}=(3, -3)$.

\section{Conclusions}
\label{sec:conclusions}

We proposed \edit{a} temporal robustness for STL specifications to account for forward and backward temporal perturbations. 
We showed the desirable properties of this new robustness notion, including soundness and the meaning of the temporal robustness in terms
of permissible forward and backward time shifts. 
We  then designed control laws for linear systems that maximize the temporal robustness objective using mixed-integer linear programming (MILP).
 Finally, we presented two case studies to illustrate how the proposed temporal robustness accounts for timing uncertainties.

\addtolength{\textheight}{-1cm}   


%

\bibliographystyle{unsrt-abbrv}
\bibliography{root}

\begin{thebibliography}{10}

\bibitem{maler2004monitoring}
O.~Maler and D.~Nickovic.
\newblock Monitoring temporal properties of continuous signals.
\newblock In {\em Formal Techniques, Modelling and Analysis of Timed and
  Fault-Tolerant Systems}, pages 152--166. Springer, 2004.

\bibitem{fainekos2009robustness}
G.~E. Fainekos and G.~J. Pappas.
\newblock Robustness of temporal logic specifications for continuous-time
  signals.
\newblock {\em Theoretical Computer Science}, 410(42):4262--4291, 2009.

\bibitem{gilpin2020smooth}
Y.~Gilpin, V.~Kurtz, and H.~Lin.
\newblock A smooth robustness measure of signal temporal logic for symbolic
  control.
\newblock {\em IEEE Control Systems Letters}, 5(1):241--246, 2020.

\bibitem{varnai2020robustness}
P.~Varnai and D.~V. Dimarogonas.
\newblock On robustness metrics for learning {STL} tasks.
\newblock In {\em 2020 American Control Conference (ACC)}, pages 5394--5399.
  IEEE, 2020.

\bibitem{raman2014model}
V.~Raman, A.~Donz{\'e}, M.~Maasoumy, R.~M. Murray, A.~Sangiovanni-Vincentelli,
  and S.~A. Seshia.
\newblock Model predictive control with signal temporal logic specifications.
\newblock In {\em 53rd IEEE Conference on Decision and Control}, pages 81--87.
  IEEE, 2014.

\bibitem{buyukkocak2021planning}
A.~T. Buyukkocak, D.~Aksaray, and Y.~Yaz{\i}c{\i}o{\u{g}}lu.
\newblock Planning of heterogeneous multi-agent systems under signal temporal
  logic specifications with integral predicates.
\newblock {\em IEEE Robotics and Automation Letters}, 6(2):1375--1382, 2021.

\bibitem{kurtz2022mixed}
V.~Kurtz and H.~Lin.
\newblock Mixed-integer programming for signal temporal logic with fewer binary
  variables.
\newblock {\em IEEE Control Systems Letters}, 6:2635--2640, 2022.

\bibitem{mehdipour2019average}
N.~Mehdipour, C.-I. Vasile, and C.~Belta.
\newblock Average-based robustness for continuous-time signal temporal logic.
\newblock In {\em 2019 IEEE 58th Conference on Decision and Control (CDC)},
  pages 5312--5317. IEEE, 2019.

\bibitem{pant2018fly}
Y.~V. Pant, H.~Abbas, R.~A. Quaye, and R.~Mangharam.
\newblock Fly-by-logic: control of multi-drone fleets with temporal logic
  objectives.
\newblock In {\em 2018 ACM/IEEE 9th International Conference on Cyber-Physical
  Systems (ICCPS)}, pages 186--197. IEEE, 2018.

\bibitem{lindemann2018control}
L.~Lindemann and D.~V. Dimarogonas.
\newblock Control barrier functions for signal temporal logic tasks.
\newblock {\em IEEE control systems letters}, 3(1):96--101, 2018.

\bibitem{charitidou2021barrier}
M.~Charitidou and D.~V. Dimarogonas.
\newblock Barrier function-based model predictive control under signal temporal
  logic specifications.
\newblock In {\em European Control Conference, Rotterdam, the Netherlands,
  accepted}, 2021.

\bibitem{cai2022overcoming}
M.~Cai, E.~Aasi, C.~Belta, and C.-I. Vasile.
\newblock Overcoming exploration: Deep reinforcement learning in complex
  environments from temporal logic specifications.
\newblock {\em arXiv preprint arXiv:2201.12231}, 2022.

\bibitem{Donze10STLRob}
A.~Donz\'e and O.~Maler.
\newblock Robust satisfaction of temporal logic over real-valued signals.
\newblock In {\em Proceedings of the International Conference on Formal
  Modeling and Analysis of Timed Systems}, 2010.

\bibitem{rodionova2021time}
A.~Rodionova, L.~Lindemann, M.~Morari, and G.~J. Pappas.
\newblock Time-robust control for {STL} specifications.
\newblock In {\em 2021 60th IEEE Conference on Decision and Control (CDC)},
  pages 572--579, 2021.

\bibitem{rodionovatcs22}
A.~Rodionova, L.~Lindemann, M.~Morari, and G.~J. Pappas.
\newblock Temporal robustness of temporal logic specifications: Analysis and
  control design.
\newblock {\em arXiv preprint arXiv:2203.15661}, 2022.

\bibitem{akazaki2015time}
T.~Akazaki and I.~Hasuo.
\newblock Time robustness in {MTL} and expressivity in hybrid system
  falsification.
\newblock In {\em International Conference on Computer Aided Verification},
  pages 356--374. Springer, 2015.

\bibitem{deshmukh2015quantifying}
J.~V. Deshmukh, R.~Majumdar, and V.~S. Prabhu.
\newblock Quantifying conformance using the skorokhod metric.
\newblock In {\em International Conference on Computer Aided Verification},
  pages 234--250. Springer, 2015.

\bibitem{abbas2014formal}
H.~Abbas, H.~Mittelmann, and G.~Fainekos.
\newblock Formal property verification in a conformance testing framework.
\newblock In {\em 2014 Twelfth ACM/IEEE Conference on Formal Methods and Models
  for Codesign (MEMOCODE)}, pages 155--164. IEEE, 2014.

\bibitem{buyukkocak2022temporal}
A.~T. Buyukkocak and D.~Aksaray.
\newblock Temporal relaxation of signal temporal logic specifications for
  resilient control synthesis.
\newblock {\em arXiv preprint arXiv:2208.08384}, 2022.

\bibitem{sahin2017synchronous}
Y.~E. Sahin, P.~Nilsson, and N.~Ozay.
\newblock Synchronous and asynchronous multi-agent coordination with {cLTL+}
  constraints.
\newblock In {\em 2017 IEEE 56th Annual Conference on Decision and Control
  (CDC)}, pages 335--342. IEEE, 2017.

\bibitem{sahin2019multirobot}
Y.~E. Sahin, P.~Nilsson, and N.~Ozay.
\newblock Multirobot coordination with counting temporal logics.
\newblock {\em IEEE Transactions on Robotics}, 36(4):1189--1206, 2019.

\bibitem{lindemann2022temporal}
L.~Lindemann, A.~Rodionova, and G.~Pappas.
\newblock Temporal robustness of stochastic signals.
\newblock In {\em 25th ACM International Conference on Hybrid Systems:
  Computation and Control}, pages 1--11, 2022.

\bibitem{selvaratnam2022mitl}
D.~Selvaratnam, M.~Cantoni, J.~Davoren, and I.~Shames.
\newblock {MITL} verification under timing uncertainty.
\newblock {\em arXiv preprint arXiv:2204.10493}, 2022.

\bibitem{baras_runtime}
Z.~Lin and J.~S. Baras.
\newblock Optimization-based motion planning and runtime monitoring for robotic
  agent with space and time tolerances.
\newblock In {\em 21st IFAC World Congress}, pages 1900--1905, 2020.

\bibitem{vasile2017time}
C.-I. Vasile, D.~Aksaray, and C.~Belta.
\newblock Time window temporal logic.
\newblock {\em Theoretical Computer Science}, 691:27--54, 2017.

\bibitem{kamale2021automata}
D.~Kamale, E.~Karyofylli, and C.-I. Vasile.
\newblock Automata-based optimal planning with relaxed specifications.
\newblock In {\em 2021 IEEE/RSJ International Conference on Intelligent Robots
  and Systems (IROS)}, pages 6525--6530. IEEE, 2021.

\bibitem{penedo2020language}
F.~Penedo, C.-I. Vasile, and C.~Belta.
\newblock Language-guided sampling-based planning using temporal relaxation.
\newblock In {\em Algorithmic Foundations of Robotics XII}, pages 128--143.
  Springer, 2020.

\bibitem{chen2022stl}
H.~Chen, S.~Lin, S.~A. Smolka, and N.~Paoletti.
\newblock An {STL}-based formulation of resilience in cyber-physical systems.
\newblock {\em arXiv preprint arXiv:2205.03961}, 2022.

\bibitem{lofberg2004yalmip}
J.~Lofberg.
\newblock {YALMIP}: A toolbox for modeling and optimization in matlab.
\newblock In {\em 2004 IEEE international conference on robotics and automation
  (IEEE Cat. No. 04CH37508)}, pages 284--289. IEEE, 2004.

\bibitem{gurobi}
L.~Gurobi~Optimization.
\newblock Gurobi optimizer reference manual, 2021.

\bibitem{luukkonen2011modelling}
T.~Luukkonen.
\newblock Modelling and control of quadcopter.
\newblock {\em Independent research project in applied mathematics, Espoo},
  22:22, 2011.

\end{thebibliography}

\section*{APPENDIX}
%
%

\subsection{Proof of Theorem~\ref{thm:sat_theta}}
The proof is by induction on the
structure of $\varphi$. 
We are going to prove the item \ref{i1}. Item \ref{i2} can be proven analogously. We will also only show the  predicate case, i.e., the case when $\varphi=p$. The other operators, i.e., when $\varphi=\neg\varphi\, |\,  \varphi_1\wedge\varphi_2\, |\, \varphi_1\until_I\varphi_2$, can be done analogously to \cite[Thm. 2.1]{rodionova2021time}.

\textbf{Item 1.} We must show $\theta_p(\sstraj,t) > 0 \ \Longrightarrow\ \chi_p(\sstraj,t)= 1$.
Since we are given that $\theta_p(\sstraj,t) > 0$ and in Def.~\ref{def:temporal_rob} $\tau\geq 0$, then $\chi_p(\sstraj,t) > 0$ and thus, since $\chi\in\{\pm 1\}$, $\chi_p(\sstraj,t)=1$.
	

\subsection{Proof of Theorem~\ref{thm:shift_async}}
Let $\varphi$ be an STL formula built upon a predicate set $AP=\{p_1,\ldots,p_K\}$, $\sstraj:\TDom \rightarrow X$ be a signal and $t\in\TDom$ be a time point. We want to show that for $\forall\tau_1,\ldots,\tau_K\in\Te$, such that
$\max(|\tau_1|,\ldots,|\tau_K|) \leq |\theta_\varphi(\sstraj,t)|$, it holds that  $\chi_\varphi(\sstraj_{\bar{\tau}},t )=\chi_\varphi(\sstraj,t)$.
The proof is by induction on the
structure of $\varphi$. 

\textbf{Case} $\varphi=p_k$. Denote $|\theta_{p_k}(\sstraj,t)| =r$.  		
	Then by Def.~\ref{def:temporal_rob}, 
	$\forall \kappa\in [-r,\ r]$,  $\chi_{p_k}(\sstraj,t+\kappa)=\chi_{p_k}(\sstraj,t)$.
	We get that $\chi_{p_k}(\sstraj_{\bar{\tau}},t )=\chi_{p_k}(\sstraj,t+ \tau_k ) = \chi_{p_k}(\sstraj,t)$, if $\tau_k \in [-r,r]$, i.e., if $|\tau_k|\leq r$.
	Since we assume that $\max(|\tau_1|,\ldots,|\tau_K|) \leq r$, then $|\tau_k|\leq r$ and thus $\chi_{p_k}(\sstraj_{\bar{\tau}},t )= \chi_{p_k}(\sstraj,t)$.
	 
	\textbf{Case} $\varphi=\neg\varphi_1$. By definition, $|\theta_{\varphi_1}(\sstraj,t)| =|\theta_{\varphi}(\sstraj,t)|$. 
	We are given that $\max(|\tau_1|,\ldots,|\tau_K|) \leq |\theta_{\varphi}(\sstraj,t)|=|\theta_{\varphi_1}(\sstraj,t)|$.
	The induction hypothesis leads to $\chi_{\varphi_1}(\sstraj_{\bar{\tau}},t )=\chi_{\varphi_1}(\sstraj,t)$. Thus, $\chi_{\varphi}(\sstraj_{\bar{\tau}},t )=-\chi_{\varphi_1}(\sstraj_{\bar{\tau}},t )=-\chi_{\varphi_1}(\sstraj,t)=\chi_{\varphi}(\sstraj,t)$.	 
	
	\textbf{Case} $\varphi=\varphi_1\wedge\varphi_2$. 
	We will only show the proof for the case when $\chi_\varphi(\sstraj, t)=1$, since the case when $\chi_\varphi(\sstraj, t)=-1$ can be shown analogously. 
	Since $\chi_\varphi(\sstraj, t)=1$, we know that $\chi_{\varphi_i}(\sstraj, t)=1$ for both $i\in\{1,2\}$ and also	
	due to Thm.~\ref{thm:sat_theta}, $\theta_\varphi(\sstraj,t)\geq 0$ and $\theta_{\varphi_i}(\sstraj, t) \geq 0$.
	Denote $\theta_\varphi(\sstraj,t)=r$.  Therefore, by Def.~\ref{def:temporal_rob}, $\theta_{\varphi_i}(\sstraj, t) \geq r$ for both $i\in\{1,2\}$.
	We are given $\tau_1,\ldots,\tau_K$ such that $\max(|\tau_1|,\ldots,|\tau_K|) \leq r$.  
		Therefore since $|\theta_{\varphi_i}(\sstraj, t)| \geq r$ then by the induction hypothesis for both $i\in\{1,2\}$, for given $\tau_1,\ldots,\tau_K$ it holds that 
		$\chi_{\varphi_i}(\sstraj_{\bar{\tau}},t )=1$.
		Thus, $\chi_{\varphi}(\sstraj_{\bar{\tau}},t )= \inf(\chi_{\varphi_1}(\sstraj_{\bar{\tau}},t ),\, 
		\chi_{\varphi_2}(\sstraj_{\bar{\tau}},t )) = 1=\chi_{\varphi}(\sstraj,t )$.
		

	\textbf{Case} $\varphi=\varphi_1 \until_I \varphi_2$.
	We will again only show the proof for the case when $\chi_\varphi(\sstraj, t)=1$.
	Due to Thm.~\ref{thm:sat_theta}, $\theta_\varphi(\sstraj, t)\geq 0$.
	Denote $\theta_\varphi(\sstraj, t)=r$.
Then by Def.~\ref{def:temporal_rob}, 
$\exists t'\in t+I$, such that 
		$\theta_{\formula_2}(\sstraj,t') \geq r$ and 
		$\forall t'' \in [t,t')$, $\theta_{\formula_1}(\sstraj,t'') \geq r$. Therefore, using the induction hypothesis and Thm.~\ref{thm:sat_theta}, we get that $\exists t'\in t+I$, $\chi_{\varphi_2}(\sstraj_{\bar{\tau}},t' )=\chi_{\varphi_2}(\sstraj,t')=1$ and 
		$\forall t'' \in [t,t')$, $\chi_{\varphi_1}(\sstraj_{\bar{\tau}},t'' )=\chi_{\varphi_1}(\sstraj,t'')=1$. But then $\chi_\varphi(\sstraj_{\bar{\tau}},t ) = \sup_{t'\in t+I} \inf\left(
		\chi_{\formula_2}(\sstraj_{\bar{\tau}},t'),\
		\inf_{t'' \in [t,t')} \chi_{\formula_1}(\sstraj,t'')
		\right) = 1=
		\chi_\varphi(\sstraj,t )$.
		

\subsection{Proof of Theorem~\ref{thm:bound}}
Let $\varphi$ be an STL formula, $\sstraj:\TDom \rightarrow X$ be a signal, and $t\in\TDom$ be a time point. We want to prove that $|\theta_\varphi(\sstraj,t)| \leq
|\theta^\pm_\varphi(\sstraj,t)|$.
The proof is by induction on the
structure of $\varphi$. 

\textbf{Case} $\varphi=p$.
	From Cor.~\ref{cor:theta_pm} we know that $\theta_p(\sstraj,t) =
		\chi_p(\sstraj,t)\cdot\min\left(|\thetap_p(\sstraj,t)|,\ |\thetam_p(\sstraj,t)|\right)$. Therefore, 
	$|\theta_p(\sstraj,t)| =
	\min\left(|\thetap_p(\sstraj,t)|,\ |\thetam_p(\sstraj,t)|\right) \leq |\theta^\pm_p(\sstraj,t)|$. 
	
	\textbf{Case} $\varphi=\neg\varphi_1$. Due to Def.~\ref{def:temporal_rob} and the induction hypothesis for $\varphi_1$, $|\theta_{\neg\varphi_1}(\sstraj,t)|=|\theta_{\varphi_1}(\sstraj,t)| \leq |\theta^\pm_{\varphi_1}(\sstraj,t)|=|\theta^\pm_{\neg\varphi_1}(\sstraj,t)|$. 
	
	\textbf{Case} $\varphi=\varphi_1\wedge\varphi_2$. We will again only show the proof for the case when $\chi_\varphi(\sstraj, t)=1$.
Since $\chi_\varphi(\sstraj, t)=1$ then we know that $\chi_{\varphi_i}(\sstraj, t)=1$ for both $i\in\{1,2\}$ and also	
due to Thm.~\ref{thm:sat_theta}, $\theta_\varphi(\sstraj,t)\geq 0$ and $\theta_{\varphi_i}(\sstraj, t) \geq 0$.
By induction hypothesis, for both $i\in\{1,2\}$, $\theta_{\varphi_i}(\sstraj, t) \leq \theta^\pm_{\varphi_i}(\sstraj, t)$. By Def.~\ref{def:temporal_rob}, $\theta_\varphi(\sstraj,t)=\inf(\theta_{\varphi_1}(\sstraj,t),\ \theta_{\varphi_2}(\sstraj,t))\leq \theta_{\varphi_i}(\sstraj,t)$, for both $i\in\{1,2\}$. Thus, $\theta_\varphi(\sstraj,t) \leq \theta_{\varphi_i}(\sstraj,t) \leq \inf(\theta^\pm_{\varphi_1}(\sstraj, t),\ 
		\theta^\pm_{\varphi_2}(\sstraj, t))=\theta^\pm_\varphi(\sstraj, t)$.
	

	\textbf{Case} $\varphi=\varphi_1 \until_I \varphi_2$.
		We will again only show the proof for the case when $\chi_\varphi(\sstraj, t)=1$.
	Due to Thm.~\ref{thm:sat_theta}, $\theta_\varphi(\sstraj, t)\geq 0$.
	Denote $\theta_\varphi(\sstraj, t)=r$.
	Then by Def.~\ref{def:temporal_rob}, 
	$\exists t'\in t+I$, such that 
	$\theta_{\formula_2}(\sstraj,t') \geq r$ and 
	$\forall t'' \in [t,t')$, $\theta_{\formula_1}(\sstraj,t'') \geq r$.
	 By using the induction hypothesis together with the above, we get that, $\exists t'\in t+I$, $\theta_{\formula_2}^\pm(\sstraj,t') \geq r$
		and $\forall t'' \in [t,t')$, $\theta_{\formula_1}^\pm(\sstraj,t'') \geq r$. 
	Therefore, $\theta^\pm_\varphi(\sstraj,t) = \sup_{t'\in t+I}\inf \left(\theta_{\formula_2}^\pm(\sstraj,t'),\ 
	\inf_{t'' \in [t,t')} \theta^\pm_{\formula_1}(\sstraj,t'')\right)\geq r$.		
	


\end{document}